\RequirePackage{lineno}
\documentclass[a4paper,10pt,twocolumn,showkeys,nofootinbib]{revtex4-1}
\usepackage{graphicx,url}
\usepackage{amsmath, amssymb, setspace, color}
\begin{document}

\vspace*{-15ex}

\title{Whole-body counter surveys of over 2700 babies and small children in and around Fukushima Prefecture 33 to 49 months after the Fukushima Daiichi NPP accident}
{\author{Ryugo S. Hayano}
\thanks{Correspondence should be addressed: R. Hayano, (\url{hayano@phys.s.u-tokyo.ac.jp}).}
\affiliation{Department of Physics, The University of Tokyo, Tokyo 113-0033, Japan}
\author{Masaharu Tsubokura}
\affiliation{Division of Social Communication System for Advanced Clinical Research, 
Institute of Medical Science, 
The University of Tokyo, Tokyo 108-8639, Japan }
\author{Makoto Miyazaki}
\affiliation{Department of Radiation Health Management,
Fukushima Medical University, Fukushima 960-1295, Japan}
\author{Akihiko Ozaki, Yuki Shimada, Toshiyuki Kambe, Tsuyoshi Nemoto, Tomoyoshi Oikawa and Yukio Kanazawa}
\affiliation{Department of Radiation Protection, Minamisoma Municipal General Hospital, Minamisoma, Fukushima 975-0033, Japan}

\author{Masahiko Nihei and Yu Sakuma}
\affiliation{Hirata Radioactivity Inspection Center, Hirata Central Hospital, Hirata, Fukushima 963-8202, Japan}

\author{Hiroaki Shimmura, Junichi Akiyama and Michio Tokiwa}
\affiliation{Department of Radiation Protection, Iwaki Urological Clinic, Tokiwa Foundation, Iwaki, Fukushima, 973-8403, Japan}

\begin{abstract}
BABYSCAN, a whole body counter (WBC) for small children was developed in 2013, and units have been installed at three hospitals in Fukushima Prefecture. Between December, 2013 and March, 2015, 2707 children between the ages of 0 and 11 have been scanned, and none had detectable levels of radioactive cesium. The minimum detectable activities (MDAs) for $^{137}$Cs were $\leq 3.5$ Bq\,kg$^{-1}$ for ages 0--1, decreasing to $\leq 2$ Bq\,kg$^{-1}$ for ages 10--11. Including the $^{134}$Cs contribution, these translate to a maximum committed effective dose of $\sim 16 \mu$Sv\,y$^{-1}$ even for newborn babies, and therefore the internal exposure risks can be considered negligibly small.

Analysis of the questionnaire filled out by the parents of the scanned children regarding their families' food and water consumption revealed that the majority of children residing in the town of Miharu regularly consume local or home-grown rice and vegetables, while in Minamisoma, a majority avoid tap water and produce from Fukushima. The data show, however, no correlation between consumption of locally produced food and water and the children's body burdens.

\end{abstract}
\keywords{Fukushima Daiichi accident, radioactive cesium, whole-body counting, committed effective dose, BABYSCAN}
\maketitle

\section{Introduction}

The Fukushima Daiichi NPP accident~\cite{tanaka} contaminated the soil of densely-populated regions of Fukushima Prefecture with radioactive cesium, which poses risks of internal and external exposures to the residents. However, most of the data accumulated and disseminated so far have consistently shown that the internal contamination for the majority of residents has fortunately been so low as to be undetectable~\cite{unscear2013}. These data include, for example, whole-body-counter surveys~\cite{hayano2013,nagataki,tsubo1,tsubo2}, duplicate-diet studies~\cite{coop},  and the inspection of ``all rice in all rice bags'' harvested in Fukushima (2012-2014)~\cite{rice}. 

Nevertheless, many residents, families with small children in particular, continue to be extremely concerned about internal exposures. This is in part due to the fact the whole body counters currently being used in Fukushima, such as the FASTSCAN~\cite{FASTSCAN}, are designed for radiation workers, who are adults.  While this is suitable for measuring larger uptakes in large children, it is not optimum for measuring small children, and is not suitable for infants or children unable to stand. 

A whole body counter optimized for measuring small children, called the BABYSCAN, was developed in order to fulfill the requests of families in Fukushima~\cite{babyscan} at the end of 2013, and three units have been deployed so far. This paper reports the results of the screening of 2707 children between the ages of 0-11, during the period spanning December, 2013 to the end of March, 2015.

\section{Ethics Statement}
Ethical approval for this study was granted by the ethics committee of University of Tokyo, under authorization number 25-40-1011.  Although the original data contained household addresses, it was subsequently geocoded into latitude and longitude coordinates, and pseudo-anonymized by rounding the coordinates to $1/100$ degrees prior to data analyses. The rounded coordinates were subsequently used to obtain the air dose rate near the subject's residences, and also to group the subjects according to the municipalities where they currently live.

\section{Materials and metods}

This study was conducted in collaboration with three medical institutions in Fukushima: Hirata Central Hospital (HCH: near Koriyama city); Tokiwakai Hospital (TH: in Iwaki city); and Minamisoma Municipal General Hospital (MMGH: in Minamisoma city). These locations are shown in Fig.~\ref{fig:map}. At each hospital, a whole body counter optimized for measuring small children (BABYSCAN~\cite{babyscan}) was installed, 
which was used to measure the amount of radioactive cesium ($^{134, 137}$Cs) in the bodies of children. In addition, all of the parents of the participating children were asked to complete a questionnaire regarding their family's food and water consumption, the results of which will be discussed later. 

\begin{figure*}
\includegraphics[width=\textwidth]{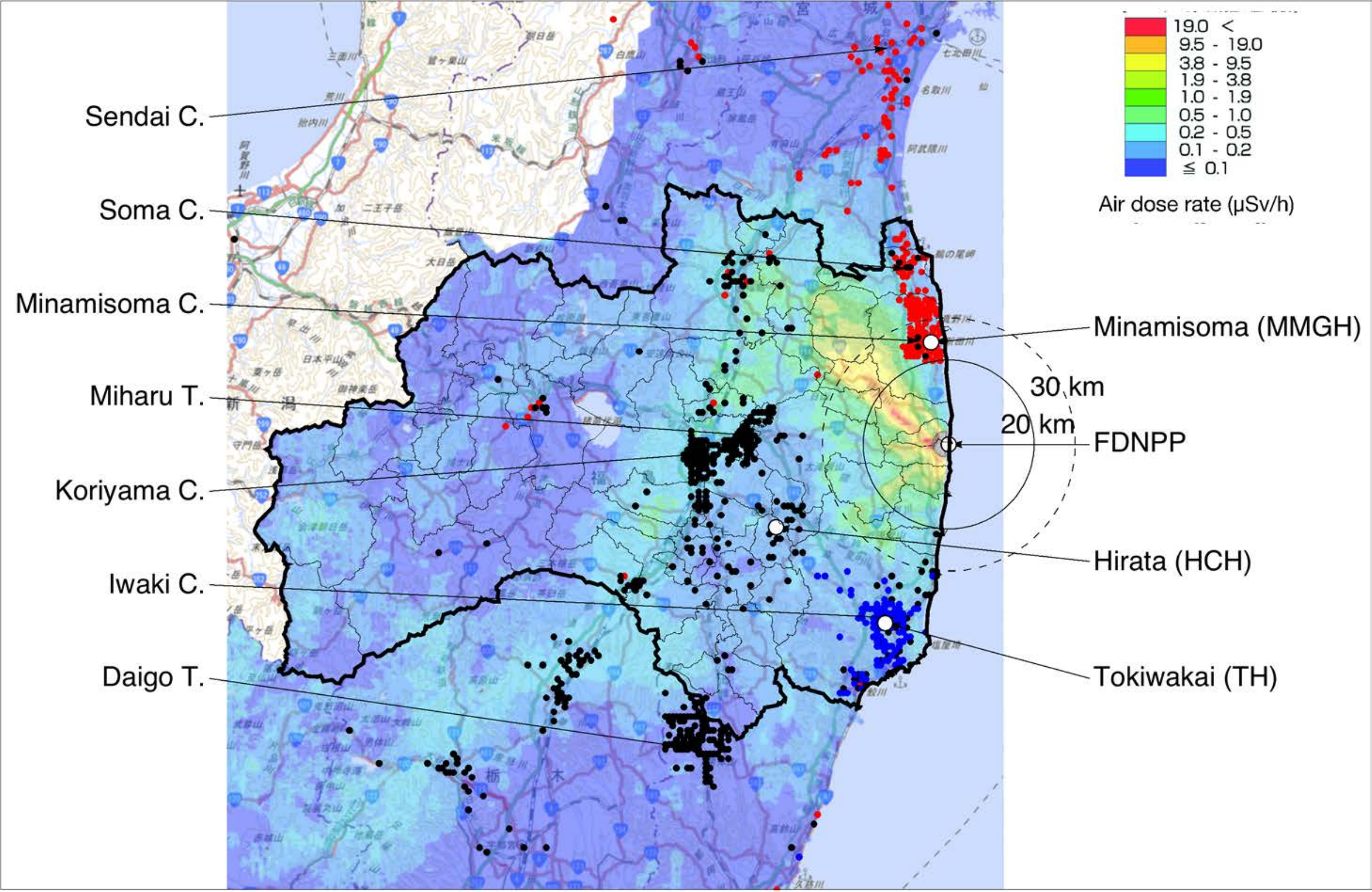}
\caption{\label{fig:map} A map showing the air-dose rate
($\mu$Sv/h) at 1m above the ground as of November 7, 2014, estimated from the 9th airborne survey conducted by the Japanese government~\cite{map}.  The boundary of Fukushima Prefecture is marked with a thick black outline. The locations of the three hospitals which participated in the present study are shown, together with the geographic locations of the subjects (black: scanned at Hirata, blue: scanned at Tokiwakai and red: scanned at Minamisoma). }
\end{figure*}

Since the technical details of BABYSCAN have already been published in Ref.~\cite{babyscan}, we here describe only the key parameters of this device.
The BABYSCAN is a whole body counter developed in 2013 for measuring small children in Fukushima (height less than 130 cm, and able to accommodate even newborn babies). During the scanning procedure, the child lies on a bed in the BABYSCAN for four minutes, during which time the gamma-rays emitted from the child's body are detected by four large sodium-iodide (NaI) detectors, which are heavily shielded against external background radiation. The minimum detectable activities (MDAs) for both $^{134,137}$Cs are less than $\sim 3$~Bq\,kg$^{-1}$ across all ages (the actual data will be presented below).

The first BABYSCAN was installed at Hirata Central Hospital in December, 2013, the second  at TH in May, 2014 and the third at MMGH in July, 2014. FASTSCAN units were already in use at each location, and the design of the BABYSCAN was intended to overcome limitations of these units when measuring small children which had become apparent during screening programs conducted from 2011 to 2013. In this study, we report the results obtained using the three BABYSCAN units from December, 2013 through the end of March, 2015. A total of 2707 children were scanned in this study period, as shown in  Fig.~\ref{fig:number}.

\begin{figure}
\includegraphics[width=\columnwidth]{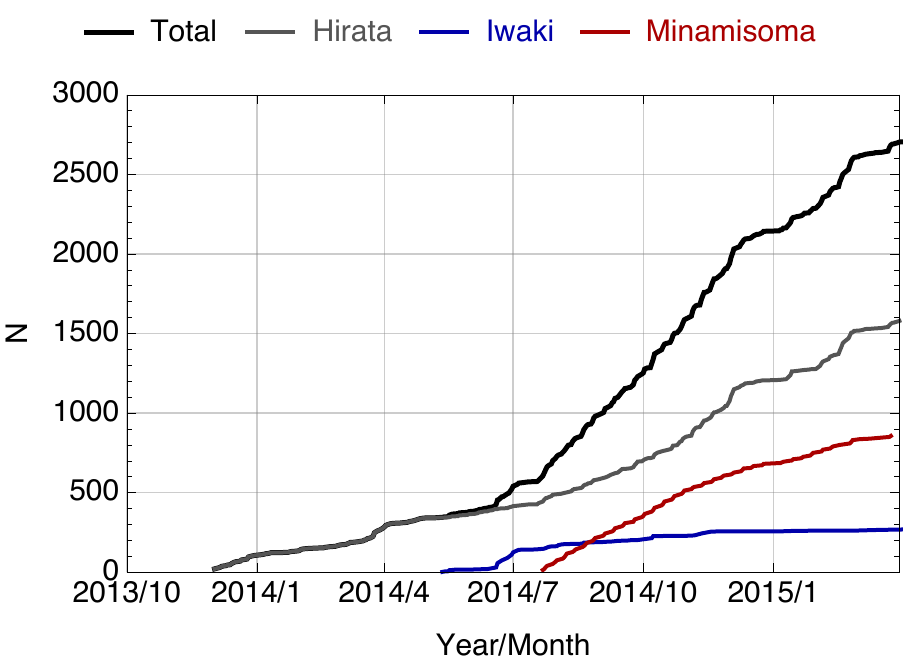}
\caption{\label{fig:number}  Cumulative numbers of subjects scanned at HCH (gray), TH (blue) and MMGH (red) during the study period. The total number (black) reached 2707 by the end of March 2015.}
\end{figure}

In Fig.~\ref{fig:map}, the  geographic distribution of the subjects scanned at Hirata Central Hospital is indicated by black dots,  that of Tokiwakai Hospital by blue dots, and of Minamisoma Municipal General Hospital by red dots. As shown, the distributions extend beyond the borders of Fukushima Prefecture. As this is related to the screening policies adopted by each hospital, some explanations are in order:

\subsection{Details of the screening programs at Hirata Central Hospital, Tokiwakai Hospital and Minamisoma Municipal General Hospital}
\subsubsection{Hirata Central Hospital}
The total number of the subjects scanned at Hirata Central Hospital was 1579. Of these, the characteristics of the participants from Miharu Town, Daigo Town, and Koriyama City are worth mentioning:

\begin{itemize}
\setlength{\topsep}{-2ex}
\setlength{\itemsep}{-1ex}
\item[-] Miharu Town - 362 subjects: In the fall of 2011, Hirata Central Hospital made an agreement with the town of Miharu to scan all the school children between the ages 6 and 15, once every year. Close to 100\% (96.2\% in the 2014 survey) of the Miharu school children have been scanned for four consecutive years, and the BABYSCAN was used for the first time in 2014 for children shorter than 130~cm, as reported recently in Ref.~\cite{miharu2014}. The present dataset includes the 2014 Miharu screening.

\item[-] Daigo Town, Ibaraki Prefecture - 431 subjects:
Hirata Central Hospital also made an agreement in 2014 to scan all children between the ages of 5 and 16 in Daigo Town, Ibaraki Prefecture,  south of Fukushima Prefecture (see Fig.~\ref{fig:map}). As in Miharu, children shorter than 130~cm were scanned using the BABYSCAN. In 2014, 1239 Daigo-town children (out of 1318 eligible, i.e., 94\%) participated in the screening, and 431 of them were screened with the BABYSCAN.

\item[-] Koriyama City - 208 subjects:
Since Hirata Central Hospital is close to Koriyama City, the most populated city in Fukushima Prefecture (population approx.\ 330,000), subjects in the  Hirata Central Hospital group include Koriyama children. Their participation in the screening was done free of charge, under an agreement with Koriyama City, and on a voluntary basis.
\end{itemize}

\subsubsection{Tokiwakai Hospital}
Tokiwakai Hospital is in Iwaki City, the second most populous city in Fukushima Prefecture (population approx.\ 325,000). Tokiwakai Hospital offers free screening using the BABYSCAN for preschool children on a voluntary basis. 
The 272 subjects scanned at Tokiwakai Hospital were mostly from Iwaki city. 

\subsubsection{Minamisoma Municipal General Hospital}
Minamisoma Municipal General Hospital, operated by the city of Minamisoma, launched a biannual WBC screening program of all school children in Minamisoma (between the ages of 6 and 15) from May 2013~\cite{minamisomaschool}. In July 2014, the city office invited children below age 6 (about 3000 in total) for free screening using the BABYSCAN. 856 children participated, of which 218 lived outside of Minamisoma city. Note that  the southern $\sim 1/3$ of the city (Odaka-ku district) was designated as within the 20-km evacuation area, and that the city was also heavily hit by the tsunami. Because both of these factors led to many evacuations, the geographical locations of the Minamisoma participants were therefore distributed far beyond the city boundary  (red dots in Fig.~\ref{fig:map}) .

\subsection{Age distribution of the subjects}
Figure~\ref{fig:age} shows the age distribution of the subjects.  As explained above, the subjects scanned at Tokiwakai Hospital and Minamisoma Municipal General Hospital were pre-school children (age $\leq 6$). At Hirata Central Hospital, school children shorter than 130~cm were also screened using the BABYSCAN, so that the age distribution extends to $\leq 11$.

\begin{figure}
\includegraphics[width=\columnwidth]{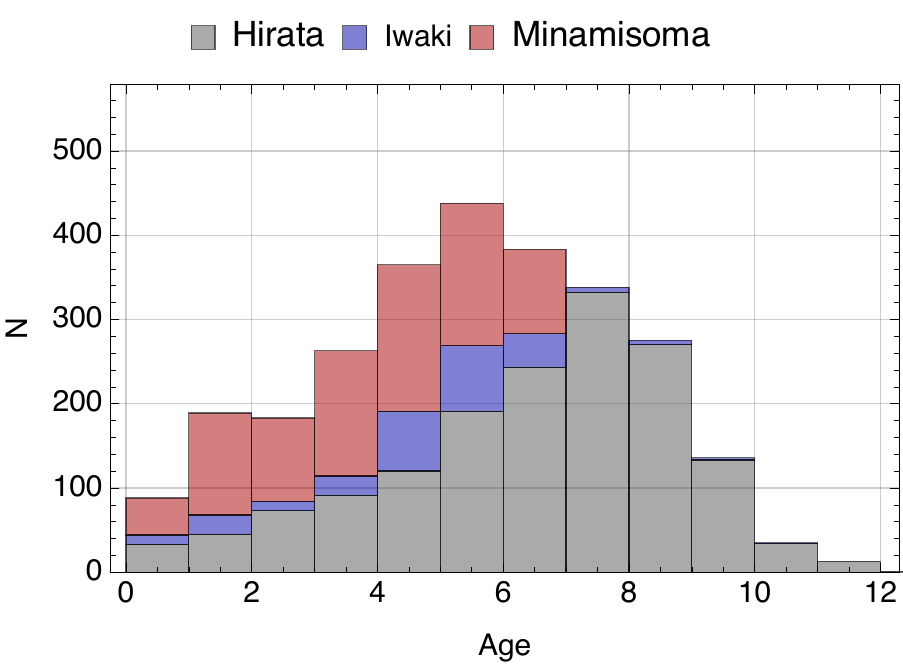}
\caption{\label{fig:age} The age distribution of the subjects; gray bars - HCH, blue bars - TH and red bars - MMGH.}
\end{figure}

\subsection{Data quality -- $^{40}$K measurement}
The quality of the WBC data can be assessed by observing the distribution of measured $^{40}$K activity in the body. The BABYSCAN was designed to ensure that the amount of $^{40}$K can be accurately quantified even for newborn babies~\cite{babyscan}. 

Figure \ref{fig:k40} shows a scatter plot of body weight (kg) versus $^{40}$K activity (Bq/body). The $^{40}$K activity was detected even for the smallest baby, confirming the high data quality of the scans.
The data points show a linear correlation between body weight and the amount of $^{40}$K in the body, with a slope of $48.8 \pm 0.3$ Bq\,kg$^{-1}$. This is consistent with the known amount of $^{40}$K in human body.

\begin{figure}
\includegraphics[width=\columnwidth]{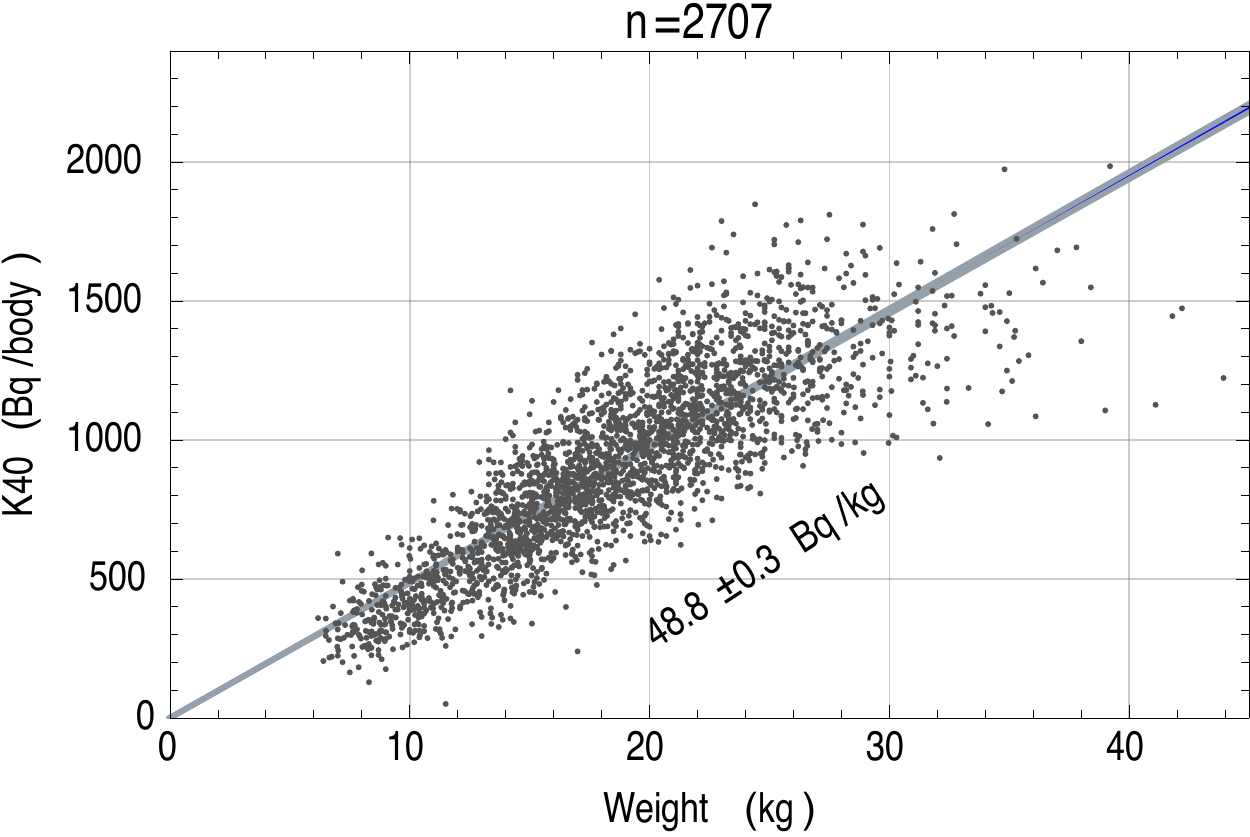}
\caption{\label{fig:k40} The distribution of the $^{40}$K activity (Bq/body) versus the body weight (kg) of the subject.}
\end{figure}

\section{Results}
During the present study period (between December 2013 and March 2015) 2707 children were scanned using  the three BABYSCAN units. No child had a detectable level of radioactive cesium in the body. To be more quantitative, a box-whisker plot of the minimum detectable activity (MDA) for $^{137}$Cs vs the subjects' ages,  Fig.~\ref{fig:mda}, shows that the MDA was $\lesssim 3.3$ Bq\,kg$^{-1}$ for all subjects.

Based on these MDAs, the estimated upper limit of ingestion is 1 (2) Bq/day, and that of the committed effective dose is $\sim 8 (16)\mu $Sv/year, including the $^{134}$Cs contribution  for 6- (0-)year olds~\cite{miharu2014}. For this reason we conclude that the concomitant health risk is negligibly small.

\begin{figure}
\includegraphics[width=\columnwidth]{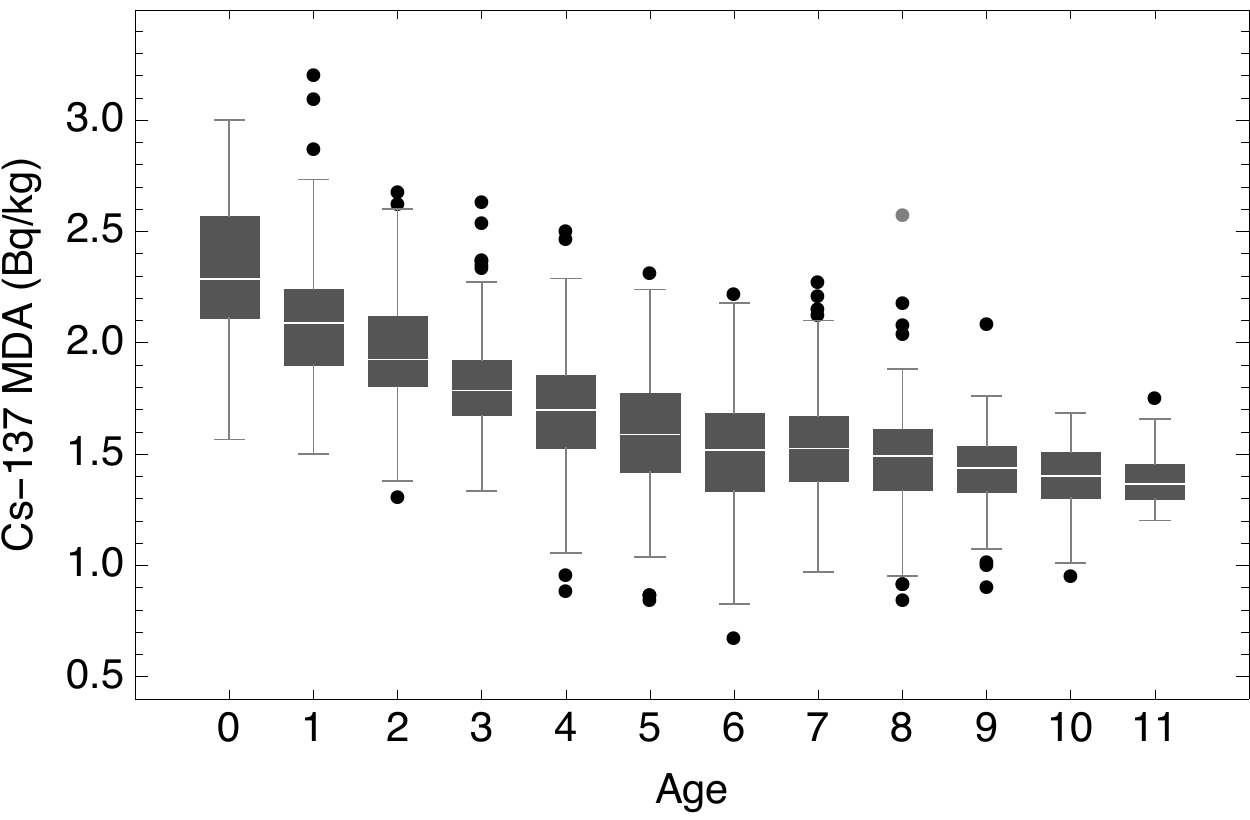}
\caption{\label{fig:mda} A box-whisker plot of the minimum detectable activity (MDA) for $^{137}$Cs vs subjects' age. The lower (upper) edge of the box represents 25- (75-) percentile. The length of the whisker is up to 1.5 times the box height, and the data beyond the whisker (outliers) are shown by dots.}
\end{figure}

\section{Discussion}
These results reconfirm that the present levels of internal exposure found in people living in currently inhabited parts of Fukushima and in surrounding prefectures can be considered low~\cite{hayano2013,nagataki,tsubo1,coop}. This is different, however, from the perception of many parents raising children in the accident-affected areas, as the  analyses of the questionnaire to the parents indicate.

In the questionnaire filled out by parents regarding their family's food and water consumption, we here focus on the answers to the following three questions:

\begin{itemize}
\setlength{\topsep}{-1ex}
\setlength{\itemsep}{-1ex}
\item[$w$] Do you avoid tap water and drink only bottled water? 
\item[$r$] Do you avoid Fukushima rice?  
\item[$v$] Do you avoid Fukushima vegetables?
\end{itemize}

\begin{table*}
\caption{\label{tab:q} Results of the questionnaire filled out by parents regarding their family's food and water consumption for the selected four regions. $N$: total number of subjects. The columns $w, r, v$ respectively show the number of subjects who drink bottled water, avoid Fukushima rice, and avoid Fukushima vegetables. The following three columns, $w\cap r$, $r\cap v$ and $v\cap w$, show the overlap counts, and the last two columns show the overlap counts of all three and its fraction (\%).}
\begin{tabular}{l|r|rrr|rrr|rr}
\hline
Region&$N$ & $w$ &$r$ & $v$ & $w\cap r$ & $r\cap v$ & $v \cap w$ & $w\cap r \cap v$ & $F(\%)$\\
\hline
Miharu &362&105&46&50&24&23&21&14&4\%\\
Minamisoma$^*$ &638&507&492&465&411&429&389&362&57\%\\
Soma$^{+}$ & 77&62&59&65&54&55&55&50&65\%\\
Daigo&431&51&33&102&5&29&19&9&1\%\\
\hline
Koriyama & 208&106&89&96&53&72&66&47&23\%\\
Iwaki&291&160&142&161&102&119&113&88&30\%\\
\hline
\end{tabular}
{\small
{\ }\\
\begin{flushleft}
*) The children whose permanent domicile is in Minamisoma city, and the current address is also in Minamisoma city.\\
+) The children whose permanent domicile is in Minamisoma, while the current address is  in Soma city.
\end{flushleft}
}
\end{table*}

Table~\ref{tab:q} summarizes the result for Miharu Town, Minamisoma City, Soma City, and Daigo Town. The Miharu and Daigo results are sampling-bias free, considering the high coverage (nearly 100\%) of the target group. The Miamisoma and Soma results may exhibit sampling bias, since the screening turnout was about 1/3, but should have less bias compared with the results from other regions, such as Koriyama and Iwaki, where participation in the screening was done on a voluntary basis (shown in the last two rows).

\begin{figure*}
\begin{minipage}{0.32\textwidth}
\includegraphics[width=0.7\columnwidth]{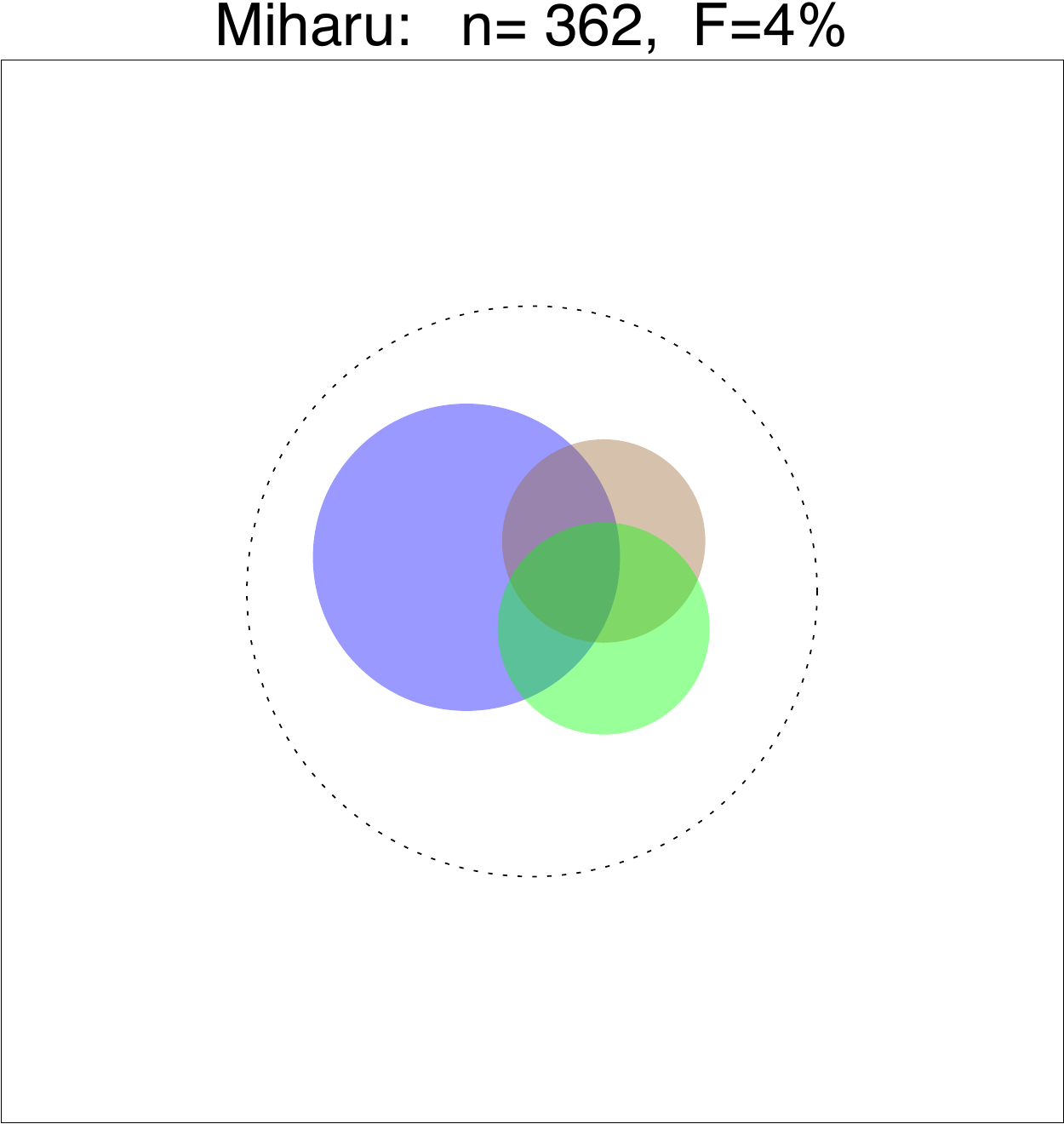}
\end{minipage}
\begin{minipage}{0.32\textwidth}
\includegraphics[width=0.7\columnwidth]{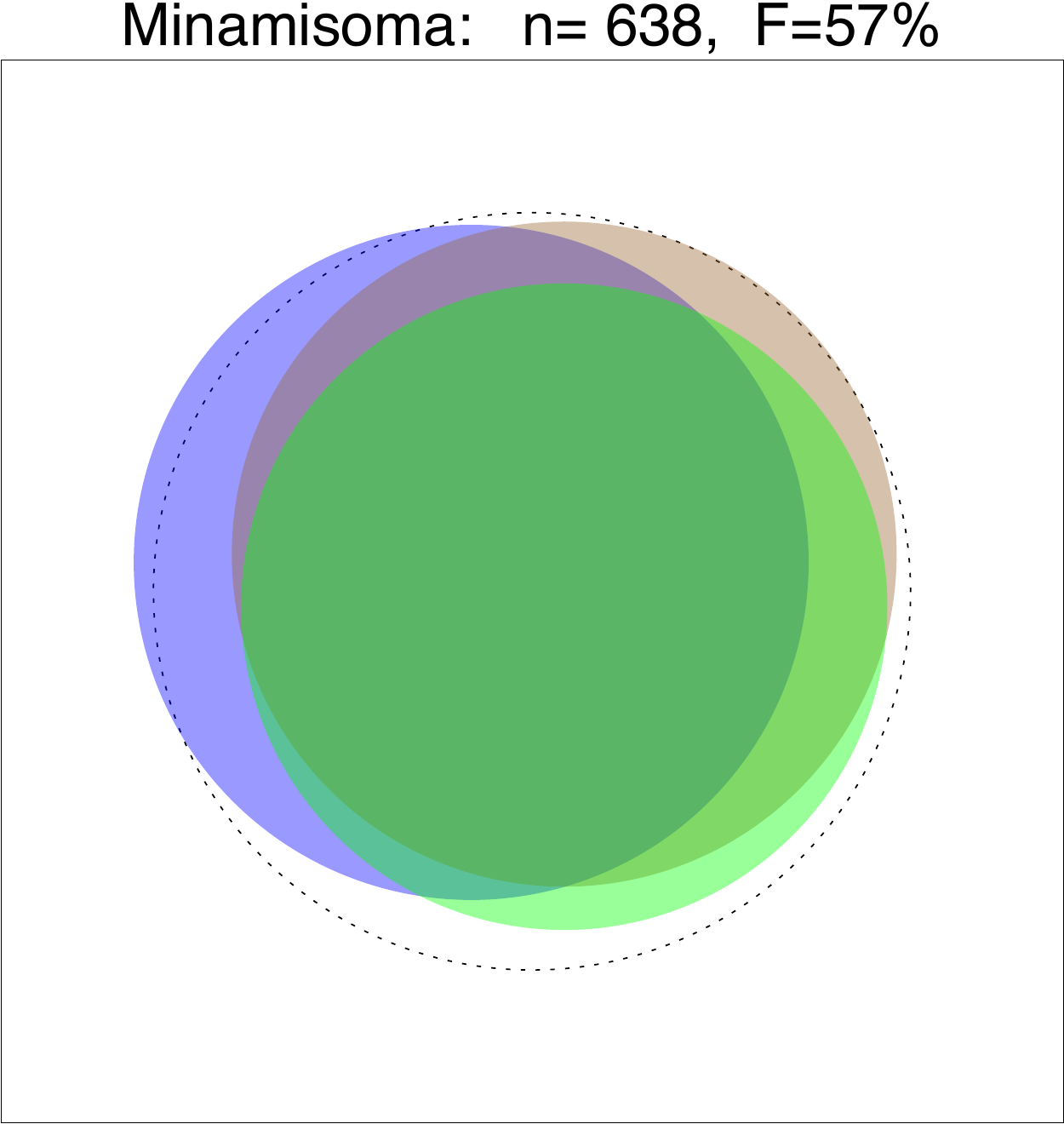}
\end{minipage}
\begin{minipage}{0.32\textwidth}
\includegraphics[width=0.7\columnwidth]{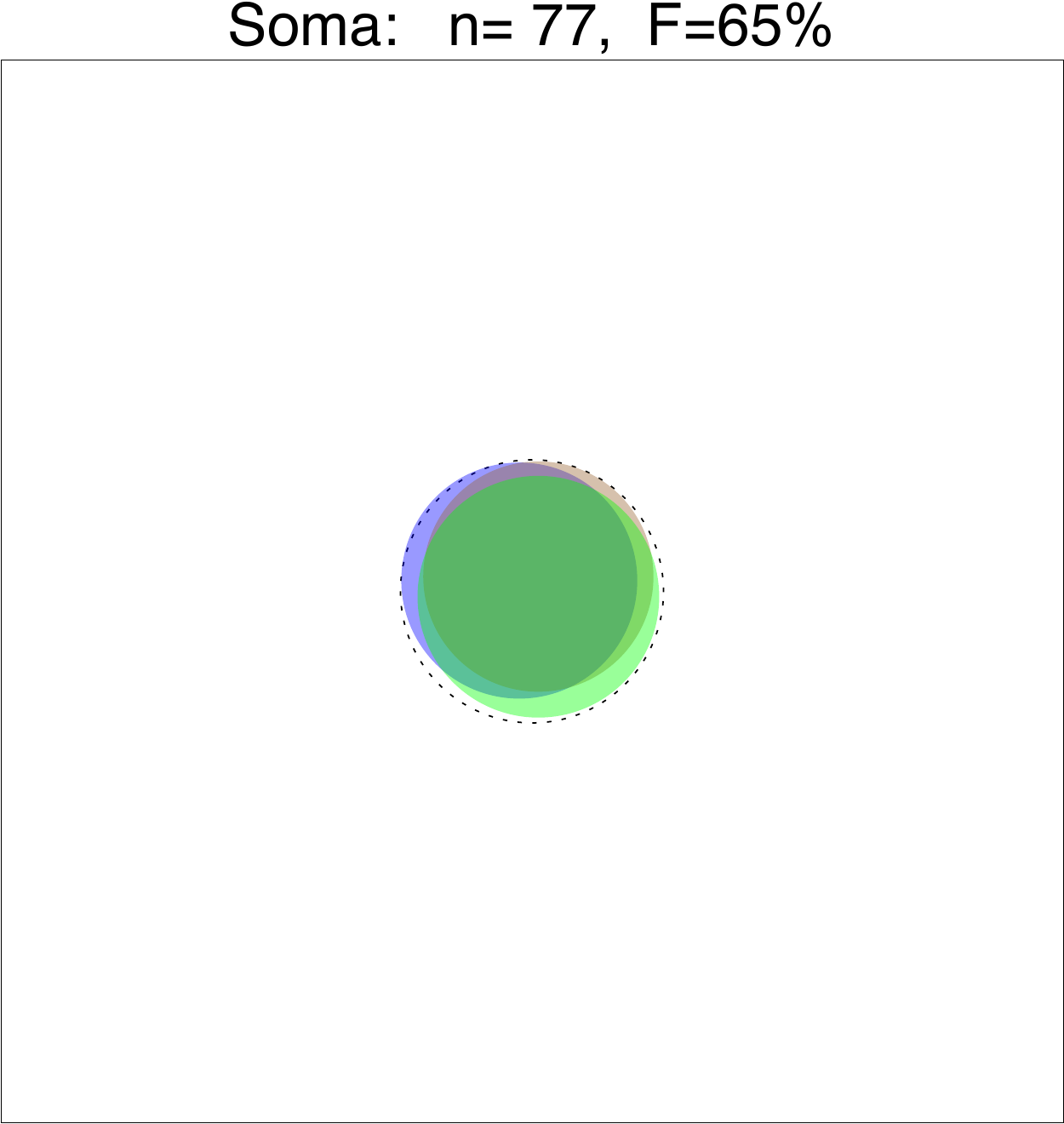}
\end{minipage}

\vspace*{2ex}

\begin{minipage}{0.32\textwidth}
\includegraphics[width=0.7\columnwidth]{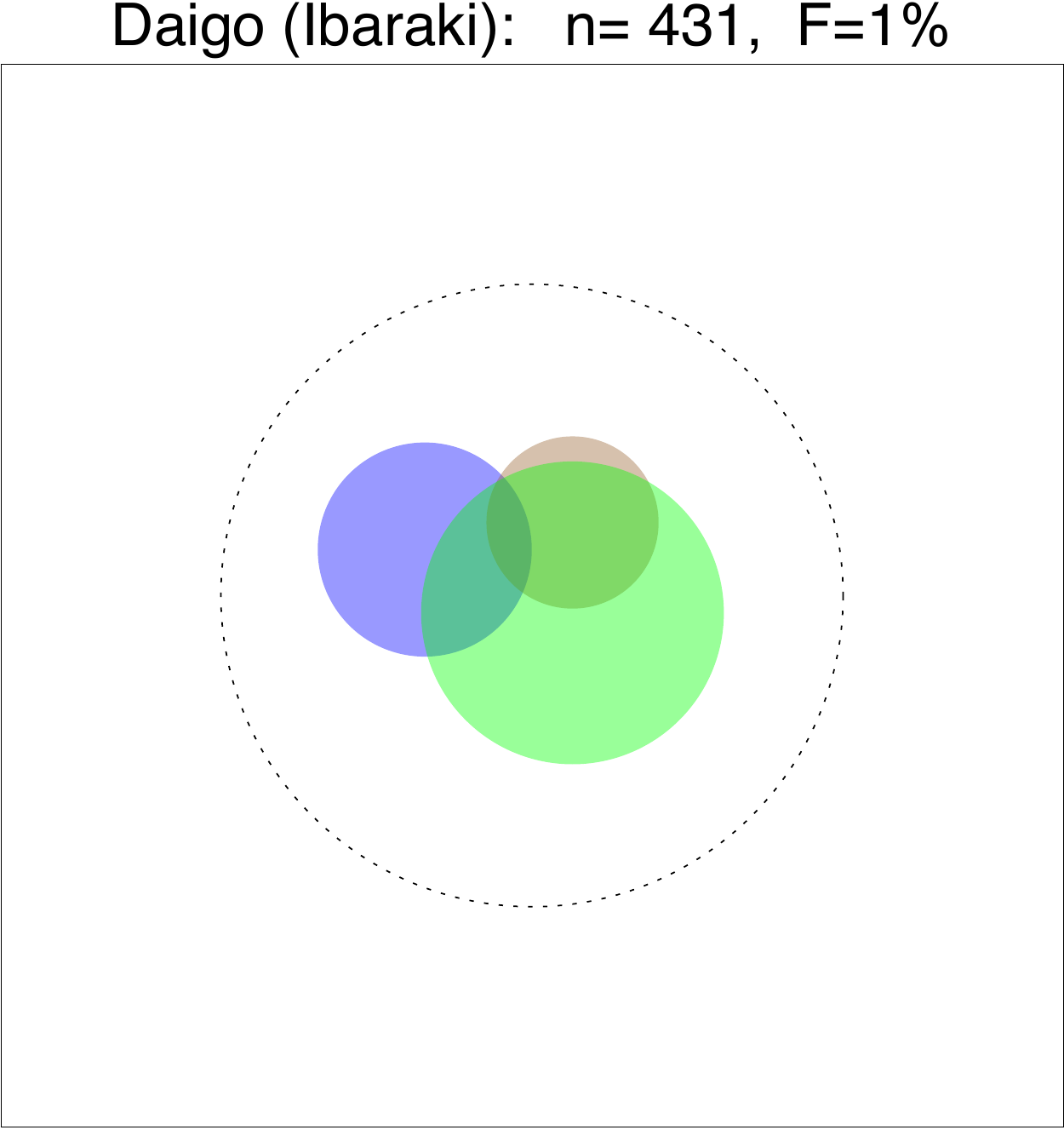}
\end{minipage}
\begin{minipage}{0.32\textwidth}
\includegraphics[width=0.7\columnwidth]{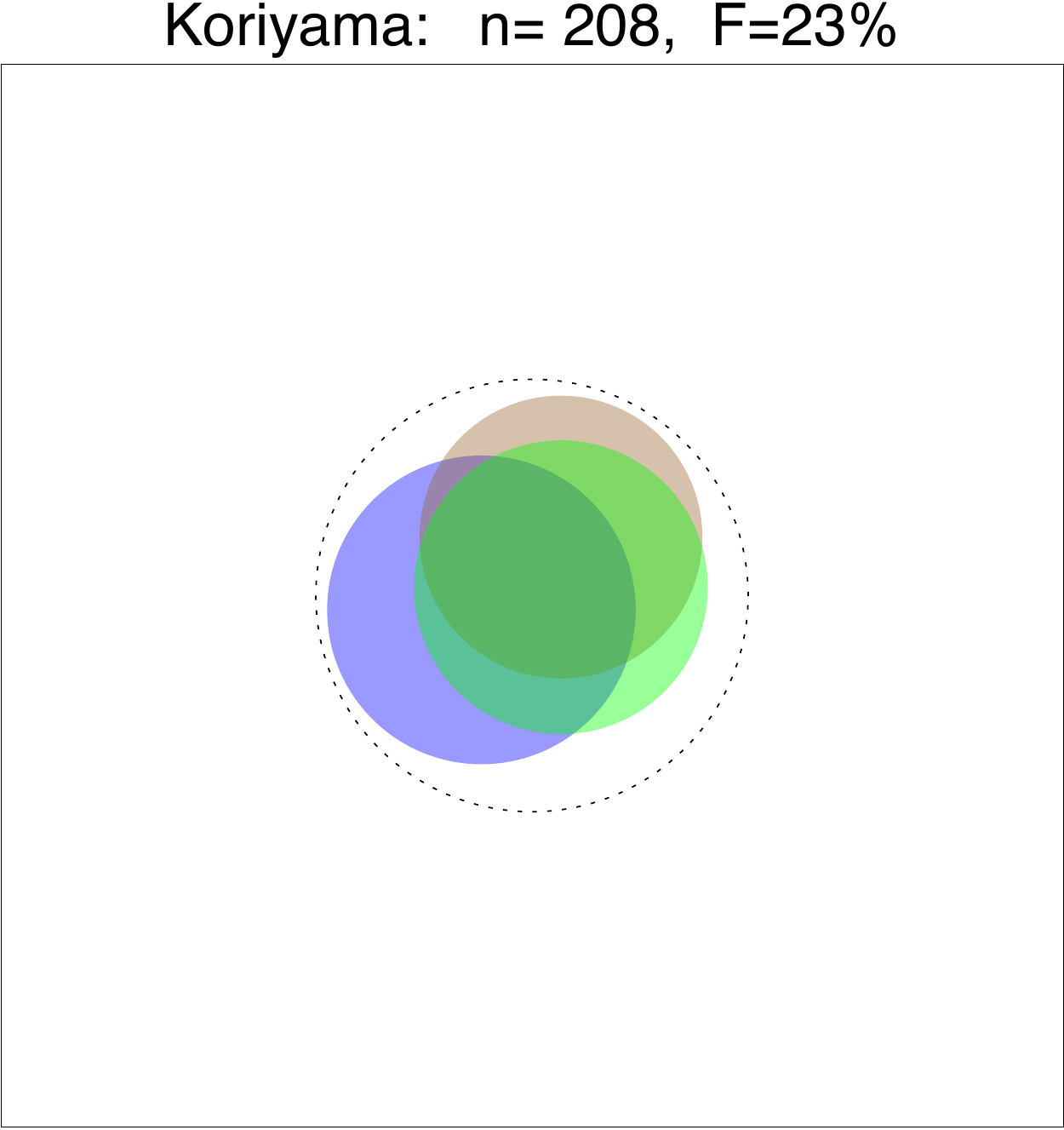}
\end{minipage}
\begin{minipage}{0.32\textwidth}
\includegraphics[width=0.7\columnwidth]{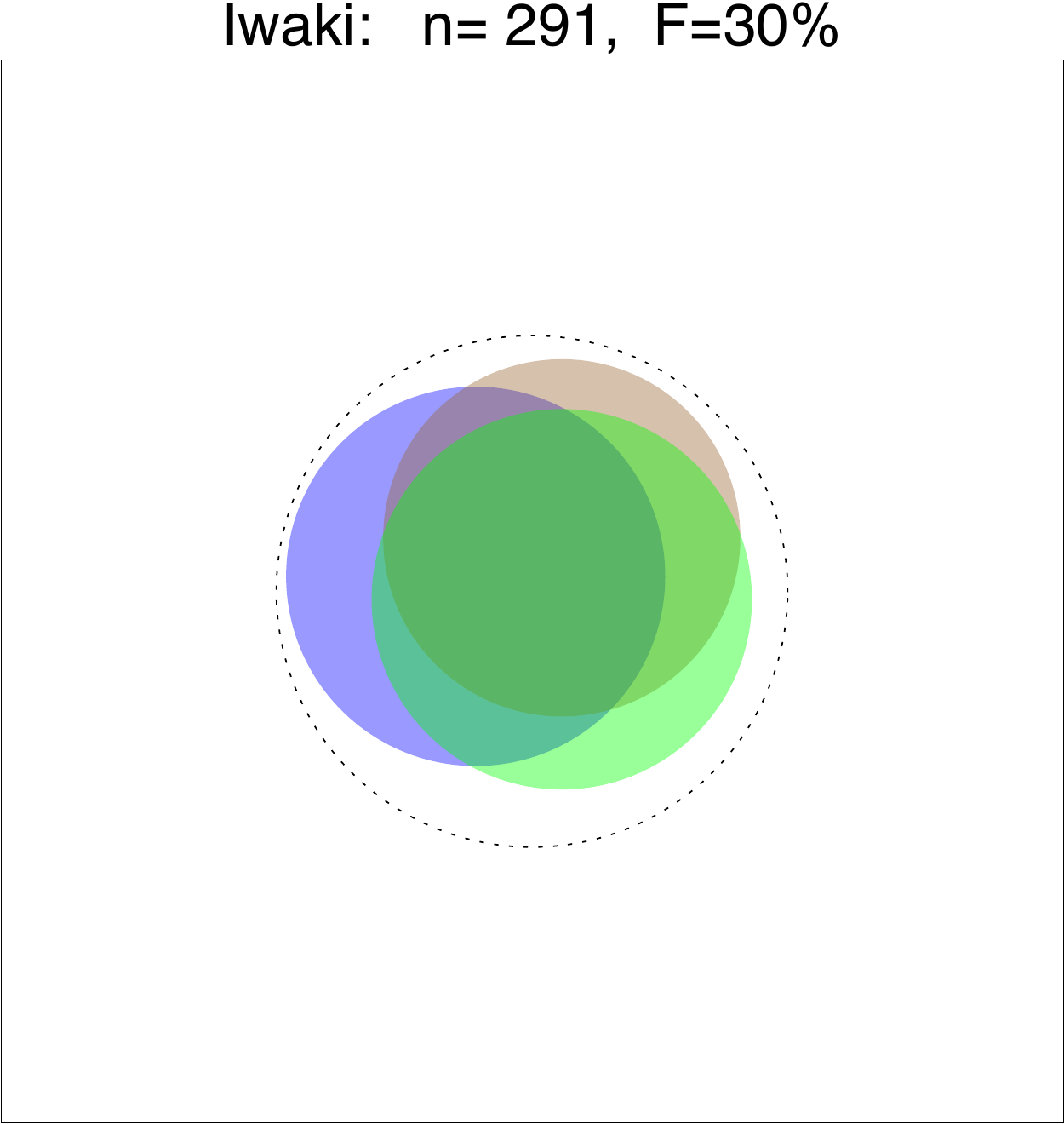}
\end{minipage}
\caption{\label{fig:q} A graphical representation of the data listed in Table \ref{tab:q}. The area of each circle is proportional to the numbers in the table. From top left to bottom right, results of Miharu Town, Minamisoma City, Soma City and Daigo Town, Koriyama City and Iwaki City. The water ($w$) results are shown in blue, rice ($r$) in red and vegetables ($v$) in green. The area of the dotted circle is proportional to $N$. }
\end{figure*}

The results are graphically presented in Fig.~\ref{fig:q}, in which the number of table entries are drawn proportional to the area of the circles ($w$: blue, $r$: red and $v$: green). The overlap counts are reflected in the overlap of the circles.

As was previously discussed in Ref.~\cite{miharu2014}, a majority of the families living in Miharu town do not avoid Fukushima (local) produce. The percentage $F$ of the subjects who drink bottled water, avoid Fukushima rice, and avoid Fukushima vegetable is 4\%. The lack of correlation with the childrens' body burdens when compared to results from other towns, however, shows that consuming these foodstuffs does not currently increase the risk of internal exposure in children.

In Daigo Town, $F$ is 1\%, but the percentage of the subjects who avoid Fukushima vegetables is higher than in Miharu (24\% vs 14\%).

The risk perceptions of Minamisoma citizens are quite different: 57 (65)\% of the Minamisoma children who currently live in Minamisoma (Soma) drink bottled water, avoid Fukushima rice, and avoid Fukushima vegetables. Here, choosing bottled water is more closely correlated with the avoidance of Fukushima rice and/or vegetables than in Miharu. The results from Koriyama and Iwaki lie between these results, but are also more prone to the sampling-bias effects.

It may be natural to assume that the risk perceptions of residents in a particular area are related to the documented levels of soil contamination where they live. However, as shown in Fig.~\ref{fig:dose},  the air dose rates near the residences of subjects in Miharu and in Minamisoma, as estimated from the airborne survey database published by the Japanese government~\cite{map}~\footnote{Note that the flight altitudes were about 300~m above ground, and the estimated air dose rate is the averaged value of air dose rates in a $\sim 600$~m diameter circle. See, e.g., Ref~\cite{jaea}}, are similar (Miharu: $0.26\pm 0.07 \mu$Sv/h vs Minamisoma: $0.25\pm 0.07 \mu$Sv/h), and are lower in Soma ($0.17\pm 0.04 \mu$Sv/h). 
\begin{figure*}
\includegraphics[width=0.32\textwidth]{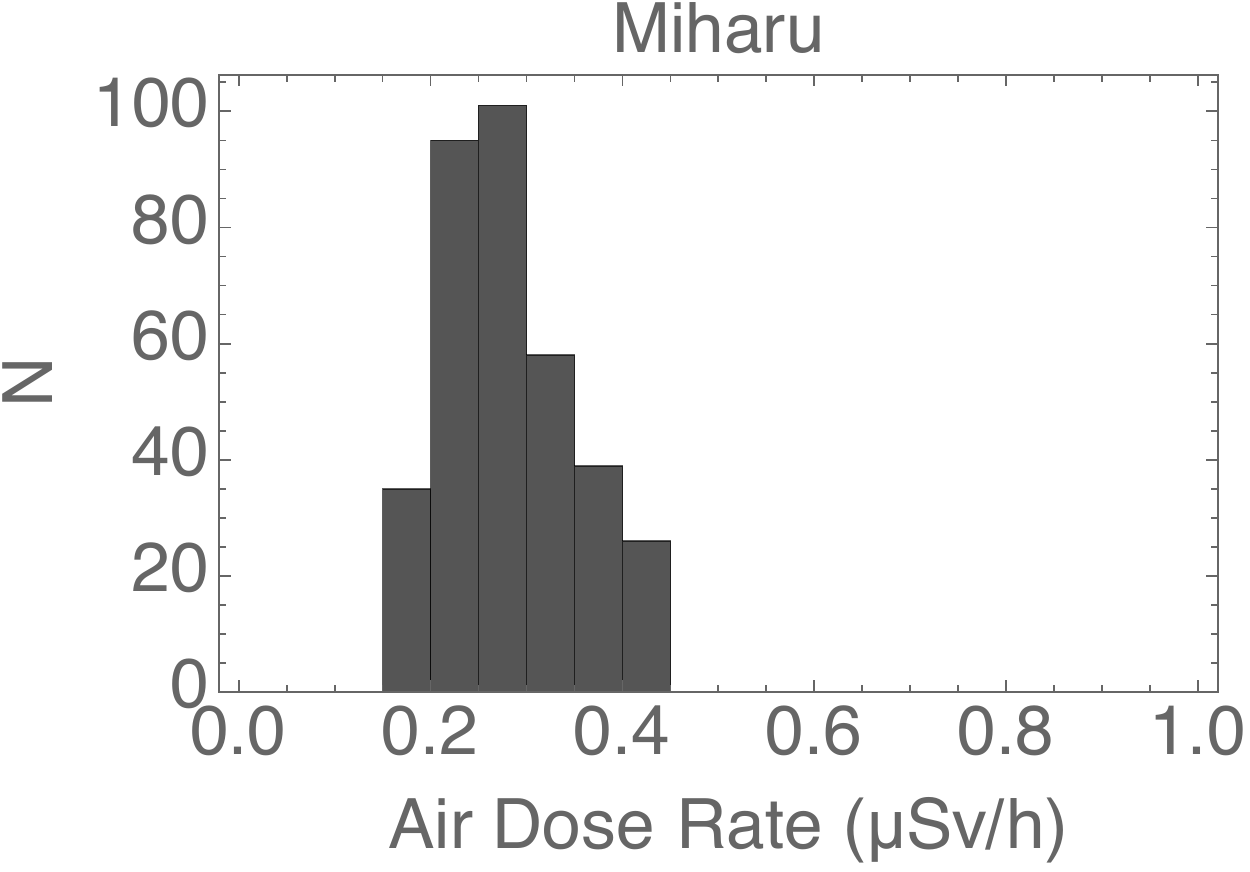}
\includegraphics[width=0.32\textwidth]{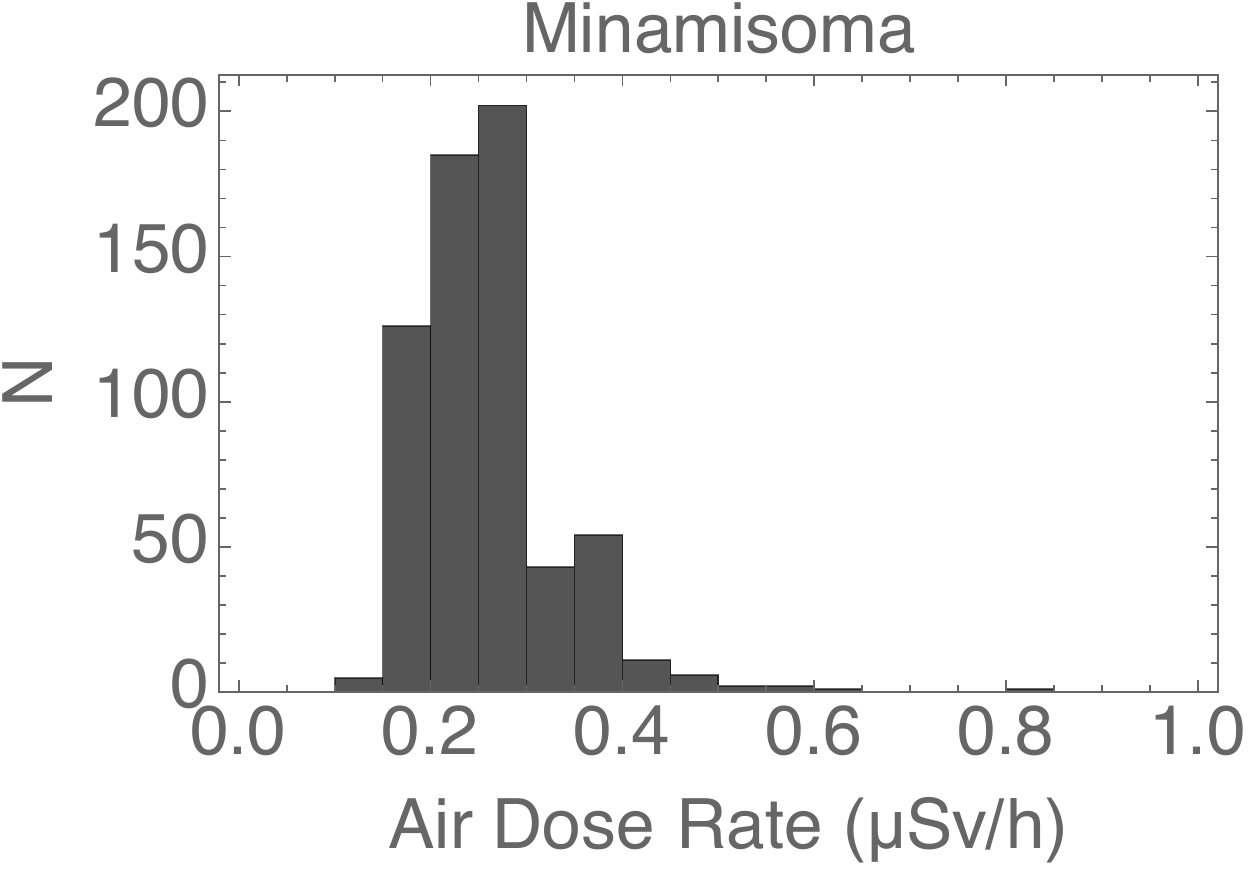}
\includegraphics[width=0.32\textwidth]{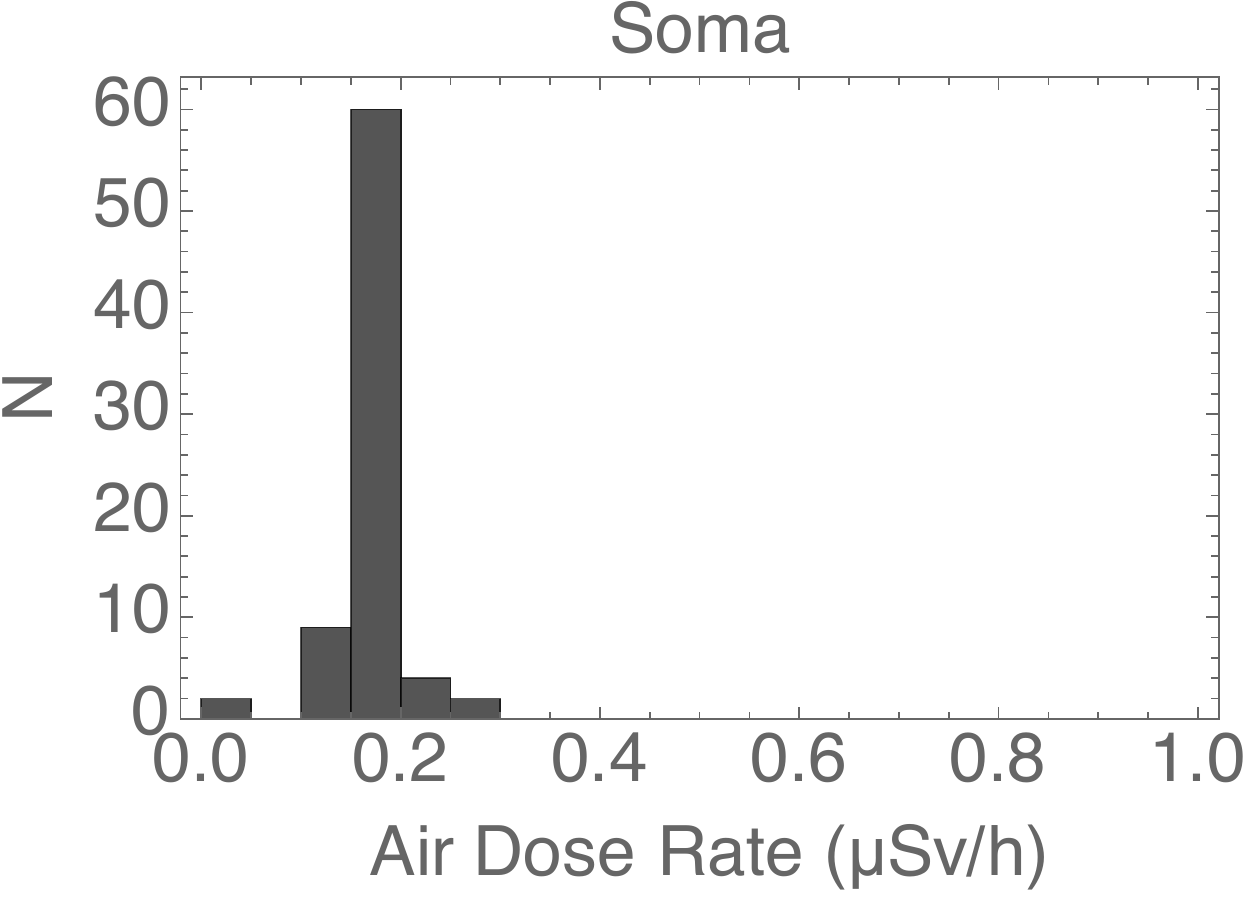}
\caption{\label{fig:dose} The distribution of the air dose rate ($\mu$Sv/h) near the residences of the subjects, estimated from the airborne survey database~\cite{map}. From left to right, Miharu Town, Minamisoma City and Soma City. }
\end{figure*}

It is not the purpose of the present paper to discuss in detail the determining factors of  risk perception in Fukushima, but these observations should offer valuable information for further studies.

\section{Conclusions}

The BABYSCAN, a whole body counter (WBC) for small children, was developed in 2013, and units have been installed at three hospitals in Fukushima Prefecture. Between December, 2013 and March, 2015, 2702 children between the ages of 0 and 11 have been scanned, and none had a detectable level of radioactive cesium. The minimum detectable activities (MDAs) for $^{137}$Cs were $\leq 3.5$ Bq\,kg$^{-1}$ for ages 0--1, decreasing to $\leq 2$ Bq\,kg$^{-1}$ for ages 10-11. Including the $^{134}$Cs contribution, these translate to a maximum committed effective dose of $\sim 16 \mu$Sv\,y$^{-1}$ even for newborn babies, and therefore the internal exposure risks are considered to be negligibly small.

Analysis of the questionnaire filled out by the parents of the screened children regarding their families' food and water consumption revealed that the majority of Miharu children regularly consume local or home-grown rice and vegetables, while in Minamisoma, a majority avoid tap water and Fukushima produce. These differences however have no detectable correlation with the body burdens.

\begin{acknowledgements}
We thank staff members at the three hospitals for the arrangement of this screening; especially Masatsugu Tanaki, Kikugoro Sakaihara, and Tatsuo Hanai, Minamisoma Municipal General Hospital; Miho Senzaki, Miki Abe, and Megumi Murakami, Hirata Radioactivity Inspection Center; Manabu Suzuki, and Ayatomo Kambara, Tokiwa Foundation. We also thank the continuing  support of the employees of local municipalities in Fukushima.
\end{acknowledgements}

\end{document}